\documentclass[a4paper,fleqn]{cas-sc}

\usepackage{geometry}
\RequirePackage[colorlinks]{hyperref}
\colorlet{scolor}{black}
\colorlet{hscolor}{DarkSlateGrey}

\usepackage[english]{babel}
\usepackage[round]{natbib}
\usepackage{url}
\usepackage{float}

\usepackage[shortlabels]{enumitem}

\def\tsc#1{\csdef{#1}{\textsc{\lowercase{#1}}\xspace}}
\tsc{WGM}
\tsc{QE}
\tsc{EP}
\tsc{PMS}
\tsc{BEC}
\tsc{DE}

\begin{document}
\let\WriteBookmarks\relax
\def\floatpagepagefraction{1}
\def\textpagefraction{.001}

\title [mode = title]{Finding the instrumental variables of household registration: A discussion of the impact of China's household registration system on the citizenship of the migrant population}

\tnotemark[1]
\tnotetext[1]{The authors thank the Migrant Population Service Center, National Health Commission P. R. China for data support.}

\author[1]{Jingwen Tan}[orcid=0000-0002-3452-959X]
\cormark[2]
\ead{tjw@henu.edu.cn}
\credit{Conceptualization of this study, Methodology, Software}
\affiliation[1]{organization={School of Economics, Henan University},
    addressline={Jinming Avenue}, 
    city={Kaifeng},
    postcode={475004}, 
    country={China}}
\fnmark[1]

\author[1]{Shixi Kang}[orcid=0000-0003-4847-5171]
\cormark[1]
\ead{ksx@henu.edu.cn}
\credit{Conceptualization of this study, Methodology, Software}
\fnmark[1]

\cortext[cor1]{Corresponding author}
\cortext[cor2]{Principal Corresponding author}
\fntext[fn1]{These authors contributed equally to this work.}


\begin{abstract}
Due to the specificity of China's dualistic household registration system and the differences in the rights and interests attached to it, household registration is prevalent as a control variable in the empirical evidence. In the context of family planning policies, this paper proposes to use family size and number of children as instrumental variables for household registration, and discusses qualitatively and statistically verifies their relevance and exogeneity, while empirically analysing the impact of the household registration system on the citizenship of the mobile population. After controlling for city, individual control variables and fixed effects, the following conclusions are drawn: family size and number of children pass the over-identification test when used as instrumental variables for household registration; non-agricultural households have about 20.2\% lower settlement intentions and 7.28\% lower employment levels in inflow cities than agricultural households; the mechanism of the effect of the nature of household registration on employment still holds for the non-mobile population group.
\end{abstract}
\maketitle

\section{Introduction}
China has experienced nearly three decades of rapid economic growth since reform and opening up in the 1980s resolved the institutional barriers to labour migration.2019 China's economic output exceeded 90 trillion yuan, an economic excellence closely associated with the massive movement of the rural population to the urban industrial sector\citep{Li2019}. By 2020, the urbanisation rate of China's resident population exceeds 60\%, with 13.52 million new jobs in cities\citep{Li2020}. The massive influx of cheap labour provides a huge impetus for local economic growth and promotes the rapid development of regional manufacturing industries. However, in recent years a structural labour shortage in the region has become increasingly common and the surplus rural labour force is drying up\citep{Fang2007,Garnaut2006}.

An emerging view is that this phenomenon is simply a manifestation of a discriminatory and segmented labour market that has evolved as a result of restrictions on population mobility\citep{Garnaut2010}. Prior to 2015, China had a dual household registration system across the country. The binary hukou system simply divided the population into agricultural and non-agricultural hukou, and the vast differences in entitlements attached to the two have had a profound impact on the regional allocation of labour endowments in China. The hukou system creates institutional barriers to the free movement of labour between provinces and cities and between urban and rural areas. For example, rural populations do not have access to many urban public services, including education and healthcare\citep{Afridi2015}, and a large number of rural children and elderly people are left behind in their hometowns and become surplus labour. For example, once they decide to move to an urban household registration, they have to give up ownership of their rural homes. With China's rapid urbanisation, the value of rural land on the urban fringe has increased significantly, and agricultural hukou, which is directly related to land rights, has acquired a higher monetary value, thus making rural migrants more inclined to retain their original hukou\citep{Chen2016}. These barriers have led to an influx of labour into cities in the form of a 'floating population', which is hardly a long-term human resource.

Starting from the institutional context of China's dualistic household registration policy, this paper attempts to estimate the impact of the nature of household registration on the citizenship of the floating population. Considering the possibility of natural persons changing their household registration status for external reasons, this paper attempts to find instrumental variables of household registration to address the potential biased estimation problem. At the same time, the paper also explores the outcomes of the reform of the household registration system in terms of achieving the citizenship of the migrant population and facilitating the allocation of labour resource endowments.

\section{Review of the literature}
Citizenship is generally defined as the process and phenomenon by which migrant workers are transformed into citizens by overcoming numerous barriers\citep{Chen2013}. There is a consensus in the academic community regarding the incentive effect of the reform of the household registration system on the inflow of people to cities. For example, \cite{Ito2008} finds through a general equilibrium model that the abolition of the household registration system will provide incentives for more rural people to flow to cities, thereby reducing the urban-rural gap. Such a reduction in the gap would also reduce potential social conflicts between the urban population and rural migrants, thus reducing the cost of maintaining social stability for the government\citep{Liu2010}. In addition, reform of the hukou system will bring social benefits including capital accumulation and spillover effects\citep{Glaeser2008}, and \cite{Song2021} simulation of the abolition of the hukou system suggests that the redistribution of labour resources brought about by the disappearance of mobility barriers will lead to a significant increase in GDP and greater economic benefits due to the optimisation of both urban and rural labour markets. The results of the simulation show that the redistribution of labour resources brought about by the removal of the household registration system will lead to a significant increase in GDP, with greater economic benefits arising from the optimisation of both urban and rural labour markets.

Regarding the mechanism of the negative effect of the household registration system on the citizenship of the migrant population, academic research has focused on two main aspects, namely the willingness to stay and the employment situation. First, the household registration system deprives migrant workers of some of the public services to which they are entitled, thereby reducing this group's willingness to stay. For example, the lack of educational benefits for the children of rural migrants, who have difficulty attending local schools, changes the migration patterns of the mobile population at different ages and makes a significant proportion more inclined to stay in the countryside, thus becoming surplus labour\citep{Golley2011}. For example, rural migrants who lack urban hukou have limited access to social insurance and health services, while inter-provincial migrants who lack hukou in the inflow area will be forced to return to the outflow area due to the lack of common access to a significant proportion of public services (e.g. health insurance)\citep{Dreger2015}. The lack of public services will also bring about an increase in the cost of living for rural migrants. They often have difficulty accessing low-cost housing provided by the government and have to bear the additional housing burden of high prices. As a major factor affecting migrants' willingness to stay, the increased cost of living will undoubtedly increase the probability of moving out of the mobile population\citep{Foote2016}.

Second, the household registration system puts migrant workers in a vulnerable position in the job market. Newly arrived migrants are employed or self-employed, and the discrimination against rural migrants in the case of employment has been well researched, for example, their incomes are often much lower than those of similarly qualified residents with a local hukou\citep{Lee2012,Ma2018}. In addition, the system can limit immigrants' self-employment; \cite{Paulson2004}find that individuals with family members working in financial institutions are more likely to start their own businesses, and that entrepreneurs are often forced to rely on their own wealth due to borrowing constraints in an imperfect financial mechanism. Migrants who lack a local household registration often find it difficult to become self-employed due to financing constraints and lack of interpersonal networks.

Most of the existing literature treats the household registration system as a general and holistic concept, ignoring the fact that it actually exists as a collection of multiple subdivisional forms such as local and non-local, agricultural and non-agricultural\citep{Song2014}. Methodologically, there is little discussion of the problem of biased estimation due to changes in the household registration of the sample. The main reason for this problem is that there is no argument or research in the academic community on the existence of instrumental variables in household registration. Proposing and arguing for instrumental variables in household registration through qualitative and quantitative methods is the main contribution of this paper.

\section{Finding the instrumental variables of household registration}
\begin{enumerate}[(1)]
\item China's Household Registration System.
\end{enumerate}
The household registration system is a household-based population management system implemented for citizens of the country settled in mainland China, which divides household attributes into agricultural and non-agricultural hukou based on geographical and family member relationships, indicating the legitimacy of natural persons living locally. This dualistic urban-rural household registration system inhibits the free movement of people between urban and rural areas and is discriminatory in terms of social welfare\citep{Cai2001}. Although the hukou management system has been phasing out the division between agricultural and non-agricultural hukou since 2014, its effects are still widespread and far-reaching.

Due to the specificity of the dualistic household registration system and the differences in the rights and interests of agricultural and non-agricultural households, the use of household registration as a control variable for demographics is common in studies on labour economics in China. However, the status of a natural person's household registration is not permanent, and agricultural and non-agricultural hukou can shift from one to the other with the social behaviour of the natural person, such as working, pursuing education, etc\citep{Yao2004,Li2003,Sun2011}. In empirical studies, when there is a possibility that the nature of the explanatory variable household registration changes with the status of the explanatory variable, potential endogeneity problems resulting from reciprocal causality may exist, causing disturbances to the unbiased assumptions of the estimates. The search for instrumental variables of household registration is therefore of considerable interest.

\begin{enumerate}[(2)]
\item Identification Strategies: The Policy Context of China's Family Planning System.
\end{enumerate}
Family planning is one of the basic national policies of China. Since 1980, the government has been promoting one child per couple nationwide, and in 1982, the Fifth Session of the National People's Congress designated family planning as a basic state policy. Family planning advocates one child per couple, hence the term "one-child policy", but the policy allows some families to have two children, such as rural couples with only one daughter or one parent who is an only child. Parents of over-born families are often subject to dismissal from public office or administrative sanctions, and the policy is more lenient in rural areas than in urban areas. These factors combine to create a tendency for rural families to have more children, i.e. holders of agricultural hukou have larger families and a larger number of children. This transmission mechanism is supported by a considerable amount of research and will not be repeated in this paper \citep{Wang2010,Zhuang2014,Yao2010}.

In 2015, China's Fifth Plenary Session of the 18th Central Committee announced the liberalisation of the second child, and the one-child policy that had been in place in mainland China for more than 30 years was officially consigned to history. Due to the lag in the impact of the family planning policy, which has lasted for more than three decades, and the specificity of intergenerational parenting behaviour, there are still significant differences in the household size and number of children between agricultural and non-agricultural households, with agricultural households being higher than non-agricultural households on these two indicators. As shown in Table 1, the differences in household size and number of children across the nature of hukou are intuitive in the grouped descriptive statistics of the 2018 China Migration Dynamics Test.

\begin{table}[htb]
\caption{Subgroup descriptive statistics. }\label{tbl2}
\begin{tabular*}{\tblwidth}{@{} LLL@{} }
\toprule
                            & Household size & Number of children \\
\midrule                            
Agricultural households     & 3.176737       & 1.313868           \\
Non-agricultural households & 2.906609       & 1.05539            \\
\bottomrule
\end{tabular*}
\end{table}

Based on the above discussion, this study proposes that an attempt can be made to use household size or the number of children as instrumental variables for agricultural and non-agricultural hukou. The instrumental variables need to satisfy exogeneity and correlation, i.e. household size or number of children is highly correlated with household registration and unrelated to the explanatory variables. This will be discussed in detail in this paper.

\begin{enumerate}[(3)]
\item Statistical validation of instrumental variables.
\end{enumerate}

How to argue statistically that household size can be used as an instrumental variable for household registration is the focus of this section of the paper. Previous studies on the validity of instrumental variables include \cite{Angrist1991} and \cite{Acemoglu2001}, who provide different arguments through qualitative discussions and quantitative analyses. \cite{Angrist1991} effectively argue that quarter of birth can be used as an instrumental variable for years of schooling by constructing different exclusionary instruments, but this method does not apply to the data in this article.  \cite{Acemoglu2001} takes into account the path-dependent nature of institutions and their changes and uses colonial mortality as an instrumental variable for institutions \cite{Acemoglu2001}. Family planning has long been a fundamental state policy in China, and its imprint as a historical institution is present in the country's demographic structure.

This paper begins with an analysis of the relevance of instrumental variables. The correlation between the instrumental variables and the explanatory variables can be verified by reporting the first-stage results of a two-stage least squares regression based on statistics. Table 2 reports the results of the interpretation of household size and number of children on the nature of household registration. Whether model (1) model (2) regresses the number of children and family size separately on the nature of household registration or model (3) regresses them jointly, both variables, family size and number of children, are significant at the 1\% level. Interpreting the economic significance of the coefficients, in the results of the separate regressions, the probability of urban residence is reduced by 2.87\% for each additional person in the mobile population's family size. The probability of urban hukou decreases by 4.75\% for each increase in the number of children of the migrant population. Both family size and the number of children are put into the regression equation and the results are significant at the 1\% level, despite the potential cointegration problem between the two variables. The coefficient on family size drops to 1.03\% and the coefficient on the number of children drops to 3.88\%. Table (4) in the subsequent section of this paper reports the results of the one-stage F-statistic, F=1365, which is significantly higher than the empirical value of 10. The household size and number of children of the migrant population are statistically correlated with the nature of their household registration.

\begin{table}[htb]
\caption{Correlation analysis of instrumental variables. }\label{tbl2}
\begin{tabular*}{\tblwidth}{@{} CCCC@{} }
\toprule
       & (1)        & (2)        & (3)        \\
       & OLS\_1     & OLS\_2     & OLS\_3     \\
\midrule      
family & -0.0287*** & -          & -0.0103*** \\
       & (0.000790) & -          & (0.00106)  \\
child  & -          & -0.0475*** & -0.0388*** \\
       & -          & (0.00102)  & (0.00137)  \\
\_cons & 0.269***   & 0.240***   & 0.261***   \\
       & (0.00281)  & (0.00178)  & (0.00283)  \\
\midrule       
N      & 152000     & 152000     & 152000     \\
R2     & 0.008      & 0.013      & 0.013      \\
\bottomrule
\end{tabular*}
\end{table}

The exogeneity of instrumental variables is equally important. If there are two or more instrumental variables for a given variable, the hypothesis of exogeneity of the instrumental variable can be tested by means of the p-statistic of the overidentification test. In the case of exactly identified cases, the specification of the conditions for the exogeneity of the instrumental variable can only be addressed by a qualitative discussion. This paper takes this reason into account by including both the number of children and household size variables together as instrumental variables for household registration. Admittedly, there is some correlation between the two variables of household size and number of children, and there may be problems of co-linearity in the regressions that affect the significance of the estimates. However, in the one-stage regressions, both were put into the regression equation for the nature of household registration at the same time and still reached the 1\% level of significance. To address the potential over-identification and heteroskedasticity problems of multiple instrumental variables, this paper will add generalised moment estimation and iterative generalised moment estimation to the two-stage least squares regression, while reporting the results of limited information maximum likelihood estimation.

The exogeneity of the instrumental variables depends simultaneously on the explanatory variables. In this paper, with reference to previous studies, the selected citizenship indicator, i.e. the explanatory variable, is: "whether the migrant population is willing to settle in the inflow area". Using this as the explanatory variable, the results of the test for overidentification of the number of children and family size as instrumental variables of household registration are:

Score chi2(1) = 1.73991 (p = 0.1872)

Where the p-test rejects the original hypothesis that at least one of the two instrumental variables is endogenous, it passes the over-identification test and proves that the number of children and family size of the mobile population sample have no direct effect on their willingness to settle in the incoming city. The hypothesis of exogeneity is met when family size and the number of children of the mobile population are used as instrumental variables for household registration.

\section{Research Design}
\begin{enumerate}[(1)]
\item Data introduction.
\end{enumerate}
This paper selects data from the 2018 China Migrants Dynamic Survey (CMDS) organised by the National Health Care Commission for empirical analysis. The survey covered 31 provinces (municipalities and autonomous regions) in mainland China, and the sample was selected from the migrant population who had stayed in the local area for more than one month. The average age of the sample is in the 15-59 years old range. A stratified, multi-stage and large-scale PPS sampling method is used to investigate in detail the development, individual characteristics, social integration and employment of China's migrant population, providing data support for social science research in the direction of labour economics and demography. Data related to urban control variables are obtained from the statistical yearbooks of various cities.

At the same time, in order to discuss whether the impact of the household registration system on the employment situation of the Migrant population is limited by the status of the Migrant population, this paper uses the questionnaires of the mobile and household population from the 15-year China Migrants Dynamic Survey data, which are estimated by the Heckman two-step method.

\begin{enumerate}[(2)]
\item Model design.
\end{enumerate}
The explained variables in this paper are dummy variables, and the binary models that can be used include Probit and Logit. According to \citep{Nunn2011}, the regression process has fewer least squares estimation assumptions and is more robust to statistically significant results than methods such as likelihood estimation, and it is still recommended to use the LPM method when the explanatory variable is a dummy variable or an ordered variable.
The basic expressions of the IVLPM model are.

\newtheorem{theorem}{Theorem}
\begin{theorem}
\begin{eqnarray}\label{101}
T_{i j}=\alpha+\beta h i_{i j}+X_{i j}^{\prime} \gamma+\varepsilon_{i j}, \varepsilon_{i j} \sim N\left(0, \quad \sigma^{2}\right)
\end{eqnarray}
\end{theorem}

\begin{theorem}
\begin{eqnarray}\label{102}
h i_{i j}=k+\beta_{1} z_{i j}+\beta_{2} u_{i j}++X_{i j}^{\prime} \gamma+\varepsilon_{i j}, \varepsilon_{i j} \sim N\left(0, \quad \sigma^{2}\right)
\end{eqnarray}
\end{theorem}

In equation (1), $T_{i j}$is the willingness of the mobile population to settle; ${hi}$is the nature of the household registration of the mobile family, and $X_{i j}^{\prime} \gamma$ is an individual, urban control variable. $\varepsilon_{i j}$ is the random disturbance term. 
In equation (2), z, u are the number of children and household size of the migrant population respectively. Based on the estimation of two-stage least squares, this paper reports the results of generalised moment estimation, iterative generalised moment estimation, and limited information great likelihood estimation simultaneously.

The level of citizenship of the migrant population is not only reflected by their willingness to settle in the inflow area, but also by their employment status in the inflow area, which objectively reflects their ability to integrate into the inflow city. Therefore, this paper further considers the employment status of the mobile population as a citizenship indicator. The employment status of the migrant population is correlated with its own demographic characteristics that reflect the quality of the labour force, such as education and gender, and direct estimation may give rise to endogeneity problems caused by sample selection bias, which is tested for robustness through propensity score matching.

The paper further speculates that the mechanism of action between the employment status of the sample and the nature of its household registration may be influenced by whether the sample is a local resident as a status. To test this conjecture, the paper uses the Heckman two-step method through the Household Registration Survey of the 2015 China Mobility Monitoring Data.

\begin{enumerate}[(3)]
\item Descriptive statistical analysis.
\end{enumerate}
The explanatory variable in this paper is the willingness to settle of the migrant population, and we choose the CMDS questionnaire "Are you willing to settle in the local area?" The core explanatory variable in this paper is the nature of household registration, in which agricultural households are assigned a value of 0 and non-agricultural households a value of 1. Agricultural to resident households are assigned a value of 0 and non-agricultural to resident households a value of 1.

The control variables in this paper include two parts: urban control variables and individual control variables. In existing studies on the citizenship of the migrant population, factors such as years of education, family size, age and insurance participation are frequent demographic control variables. Variables such as the level of urban healthcare and disposable income per capita are frequently found as urban control variables. In this paper, gender, years of education, work status, range of mobility, logarithmic total household income, and logarithmic total household housing expenditure of the mobile population are selected as individual control variables. GDP per capita, industrial structure, number of primary school teachers per 10,000 population, and number of hospital beds per 10,000 population were used as urban control variables.
Table 3 reports the descriptive statistics for all variables in the study. Approximately 30\% of all migrants have a desire to settle in the inflow area.

\begin{table}[htb]
\caption{Variable descriptive statistics. }\label{tbl2}
\begin{tabular*}{\tblwidth}{@{} CCCCCCCC@{} }
\toprule
Variable Name & Explanation of variables                                                                                   & N      & mean   & sd                         & min                        & max     \\
\midrule  
stay          & \begin{tabular}[c]{@{}c@{}}Willingness to settle \\ (1 willing; 0 unwilling)\end{tabular}                  & 132993 & 0.307  & 0.461                      & 0                          & 1       \\
\midrule  
hukou         & \begin{tabular}[c]{@{}c@{}}Nature of household \\ (1 non-agricultural; 2 agricultural)\end{tabular}        & 132993 & 0.179  & 0.384                      & 0                          & 1       \\
family        & Family size (number of persons)                                                                           & 132993 & 3.128  & 1.207                      & 1                          & 12      \\
\midrule  
child         & Number of children                                                                                         & 132993 & 1.268  & 0.897                      & 0                          & 9       \\
gender        & Gender (1 male; 0 female)                                                                                  & 132993 & 0.513  & 0.500                      & 0                          & 1       \\
edu           & Number of years of education                                                                               & 132993 & 10.38  & 3.430                      & 0                          & 19      \\
emploey       & \begin{tabular}[c]{@{}c@{}}Work status\\  (1 Working; 0 Not working)\end{tabular}                          & 132993 & 0.842  & 0.365                      & 0                          & 1       \\
scope         & \begin{tabular}[c]{@{}c@{}}Mobility\\ (0 inter-county; 1 inter-municipal; 2 inter-provincial)\end{tabular} & 132993 & 1.337  & 0.747                      & 0                          & 2       \\
\midrule  
lnincome      & Logarithmic total household income                                                                         & 132993 & 8.759  & 0.690                      & 0                          & 13.81   \\
lnhoscost     & Total logged housing expenditure                                                                           & 132993 & 4.893  & 3.100                      & 0                          & 10.82   \\
pgdp          & GDP per capita                                                                                             & 132993 & 95,208 & \multicolumn{1}{l}{39,754} & \multicolumn{1}{l}{12,656} & 191,942 \\
3gdp          & Share of tertiary sector GDP                                                                               & 132993 & 58.26  & 10.77                      & 26.54                      & 80.98   \\
pteacher      & Number of primary school teachers per 10,000 population                                                    & 132993 & 27.08  & 19.40                      & 0                          & 121.2   \\
pbed          & Number of hospital beds per 10,000 population                                                              & 132993 & 1,981  & 1,498                      & 0                          & 14,653  \\
\bottomrule
\end{tabular*}
\end{table}

\section{Empirical Process and Discussion}
\begin{enumerate}[(1)]
\item Household registration and willingness to settle among the migrant population.
\end{enumerate}

\begin{table}[width=.9\linewidth,cols=4,pos=h]
\caption{Instrumental variable regression.}\label{tbl1}
\begin{tabular*}{\tblwidth}{@{} CCCCC@{} }
\toprule
                                      & (4)             & (5)             & (6)             & (7)             \\
                                      & TSLS            & LIML            & GMM             & IGMM            \\
\midrule                                        
hukou                                 & -0.202 ***      & -0.203 ***      & -0.198 ***      & -0.199 ***      \\
                                      & (0.0474)        & (0.0475)        & (0.0473)        & (0.0473)        \\
gender                                & -0.0306 ***     & -0.0306 ***     & -0.0306 ***     & -0.0306 ***     \\
                                      & (0.00253)       & (0.00253)       & (0.00253)       & (0.00253)       \\
edu                                   & 0.0363 ***      & 0.0363 ***      & 0.0361 ***      & 0.0361 ***      \\
                                      & (0.00172)       & (0.00173)       & (0.00172)       & (0.00172)       \\
lnhoscost                             & -0.0227 ***     & -0.0227 ***     & -0.0227 ***     & -0.0227 ***     \\
                                      & (0.000597)      & (0.000598)      & (0.000597)      & (0.000597)      \\
lnincome                              & 0.0622 ***      & 0.0622 ***      & 0.0620 ***      & 0.0620 ***      \\
                                      & (0.00277)       & (0.00278)       & (0.00277)       & (0.00277)       \\
emploey                               & -0.168 ***      & -0.168 ***      & -0.168 ***      & -0.168 ***      \\
                                      & (0.00576)       & (0.00576)       & (0.00575)       & (0.00575)       \\
marry                                 & 0.170 ***       & 0.170 ***       & 0.170 ***       & 0.170 ***       \\
                                      & (0.00327)       & (0.00327)       & (0.00326)       & (0.00326)       \\
pgdp                                  & -0.00000123 *** & -0.00000123 *** & -0.00000123 *** & -0.00000123 *** \\
                                      & (4.25e-08)      & (4.25e-08)      & (4.24e-08)      & (4.24e-08)      \\
gdp                                   & 0.000673 ***    & 0.000674 ***    & 0.000664 ***    & 0.000664 ***    \\
                                      & (0.000161)      & (0.000161)      & (0.000160)      & (0.000160)      \\
pteacher                              & 0.000153 *      & 0.000153 *      & 0.000151 *      & 0.000151 *      \\
                                      & (0.0000887)     & (0.0000888)     & (0.0000886)     & (0.0000886)     \\
pbed                                  & -0.00000111     & -0.00000111     & -0.00000114     & -0.00000114     \\
                                      & (0.000000964)   & (0.000000965)   & (0.000000963)   & (0.000000963)   \\
\_cons                                & -0.370 ***      & -0.370 ***      & -0.366 ***      & -0.366 ***      \\
                                      & (0.0300)        & (0.0301)        & (0.0299)        & (0.0299)        \\
\midrule                                        
\textit{N}                            & 132993          & 132993          & 132993          & 132993          \\
R2                                    & 0.061           & 0.061           & 0.062           & 0.062           \\
\midrule  
\textit{One-stage regression results} & \multicolumn{4}{c}{\textit{F=1365, p=0.000}}                          \\
\textit{Over-identification test}     & \multicolumn{4}{c}{\textit{Score chi2(1) = 1.73991 (p = 0.1872)}}     \\
\bottomrule
\end{tabular*}
\end{table}

Table 4 presents the results of the instrumental variable regression models. Model (4) is the result of the IVLPM regression. Model (4) indicates that after controlling for individual as well as city-related variables, the regression coefficient of the nature of household registration on the willingness to settle of the mobile population is -0.202, which is significant at the 1\% level. The intention to settle in the inflow area is on average 20.2\% lower for mobile population groups with non-agricultural hukou compared to those with agricultural hukou.

According to model (5), the findings of the two-stage least squares regression and the limited information maximum likelihood estimation regression are similar, demonstrating that there is no weak instrumental variable problem in the empirical evidence. Models (6) and (7) using generalised moments estimation and iterative generalised moments estimation also find similar results to the two-stage least squares regression, with the heteroskedasticity problem of the potential disturbance term not significantly interfering with the empirical study and the estimation results being robust.

Judging from the economic significance of the model, the willingness to settle in the inflowing cities is significantly lower for mobile people holding non-agricultural hukou than for those holding agricultural hukou. The explanation that can be obtained is that the economic level of their hometowns is higher for non-agricultural hukou holders and lower for agricultural hukou holders, compared to the greater improvement in quality of life and stronger willingness to stay for agricultural hukou inflowing cities. It has been argued that the main motivation for migration is to obtain more economic rewards\citep{Cheng2020}. Push-pull theory is a classic theory in the field of population mobility\citep{Lee1966}. In the framework of this theory, there is an income gap between the urban and rural dual economy, and mobile people holding agricultural hukou are subject to greater push and pull from their hometown during their mobility, while non-agricultural mobile people are subject to less push and pull, and these differences are further reflected in their willingness to settle through their hukou.

An analysis of the impact of household registration on the citizenship of the migrant population from one indicator alone may not be sufficient to support the conclusion. This paper further analyses the effect of household registration on the employment status of the mobile population in an attempt to test the mechanism: again, the employment status of the mobile population with non-agricultural households is lower than the employment status of the mobile population with agricultural households.

\begin{enumerate}[(2)]
\item Household registration and the employment of the migrant population.
\end{enumerate}

\textbf{2.1 Stepwise regression results}

\begin{table}[htb]
\caption{Stepwise regression results. }\label{tbl1}
\begin{tabular*}{\tblwidth}{@{} CCCCCC@{} }
\toprule
                                      & (8)         & (9)         & (10)            & (11)            & (12)          \\
                                      & OLS         & OLS         & OLS             & OLS             & OLS           \\
\midrule                                        
hukou                                 & -0.0563 *** & -0.0902 *** & -0.0555 ***     & -0.0872 ***     & -0.0728 ***   \\
                                      & (0.00275)   & (0.00281)   & (0.00281)       & (0.00287)       & (0.00284)     \\
gender                                & -           & 0.167 ***   & -               & 0.169 ***       & 0.167 ***     \\
                                      & -           & (0.00188)   & -               & (0.00193)       & (0.00190)     \\
edu                                   & -           & 0.00826 *** & -               & 0.00845 ***     & 0.00846 ***   \\
                                      & -           & (0.000331)  & -               & (0.000342)      & (0.000344)    \\
lnhoscost                             & -           & 0.00630 *** & -               & 0.00621 ***     & 0.00524 ***   \\
                                      & -           & (0.000322)  & -               & (0.000332)      & (0.000333)    \\
lnincome                              & -           & 0.0613 ***  & -               & 0.0613 ***      & 0.0509 ***    \\
                                      & -           & (0.00171)   & -               & (0.00178)       & (0.00178)     \\
marry                                 & -           & -0.0812 *** & -               & -0.0781 ***     & -0.0623 ***   \\
                                      & -           & (0.00238)   & -               & (0.00249)       & (0.00249)     \\
pgdp                                  & -           & -           & 0.000000402 *** & 0.000000138 *** & 0.000000526   \\
                                      & -           & -           & (3.18e-08)      & (3.06e-08)      & (0.000000702) \\
gdp                                   & -           & -           & -0.000733 ***   & -0.00116 ***    & -0.000496     \\
                                      & -           & -           & (0.000101)      & (0.0000967)     & (0.00133)     \\
pteacher                              & -           & -           & 0.0000427       & -0.000161 **    & 0.0000966     \\
                                      & -           & -           & (0.0000723)     & (0.0000692)     & (0.00193)     \\
pbed                                  & -           & -           & 0.0000109 ***   & 0.00000701 ***  & 0.0000208     \\
                                      & -           & -           & (0.000000763)   & (0.000000727)   & (0.0000245)   \\
\_cons                                & 0.852 ***   & 0.184 ***   & 0.834 ***       & 0.224 ***       & 0.151         \\
                                      & (0.00105)   & (0.0139)    & (0.00574)       & (0.0149)        & (0.175)       \\
\midrule  
\textit{Individual control variables} & -           & $\surd$     & -               & $\surd$         & $\surd$       \\
\textit{Urban control variables}      & -           & -           & $\surd$         & $\surd$         & $\surd$       \\
\textit{urban fixed effect}           & -           & -           & -               & -               & $\surd$       \\
\midrule  
\textit{N}                            & 140240      & 140150      & 133080          & 132993          & 132993        \\
R2                                    & 0.004       & 0.091       & 0.007           & 0.093           & 0.131         \\
\bottomrule
\end{tabular*}
\end{table}

Table 5 presents the results of the stepwise regression model. Model (8) indicates that the regression coefficient of -0.0563 for household registration on the employment of the mobile population, conditional on the inclusion of other control variables, is significant at 1\% level, indicating that the mobile population with non-agricultural household registration is at a lower level of employment relative to the mobile population with agricultural household registration. Model (11), which includes both urban and individual control variables, has a slightly higher regression coefficient (-0.0872) compared to the regression coefficient when no control variables are included. Model (12) adds urban fixed effects to the regression controlling for both urban and individual control variables. The regression coefficient is slightly lower (-0.0728) compared to the regression coefficient without controlling for fixed effects.

In previous studies, mobile people with non-agricultural hukou were more likely to enter higher-income jobs than those with agricultural hukou, but this mechanism was not significant among lower and middle-income groups\citep{litianc2018}. In terms of the employment structure of the labour force, most regions of the country offer relatively basic jobs. At the same time, the proportion of self-employment is higher among labour force groups holding agricultural hukou\citep{Sun2017}. This may explain why mobile people holding non-agricultural hukou do not perform as well as those with agricultural hukou in terms of inflow to urban employment.

\textbf{2.2 Propensity score matching}

\begin{table}[htb]
\caption{Propensity score matching. }\label{tbl1}
\begin{tabular*}{\tblwidth}{@{} CCCCCCCCC@{} }
\toprule
\multirow{2}{*}{Matching Status} & \multicolumn{2}{c}{1:1 Matching} & \multicolumn{2}{c}{1:2 match} & \multicolumn{2}{c}{radius matching} & \multicolumn{2}{c}{nuclear matching} \\
                                 & ATT              & t             & ATT             & t           & ATT                & t              & ATT                & t               \\
\midrule                                  
pre-match                        & -.0550***        & -21.25        & -.0550***       & -21.25      & -.0550***          & -21.25         & -.0550***          & -21.25          \\
post-match                       & -.0678***        & -16.79        & -.0657***       & -17.32      & -.0694***          & -16.99         & -.0715***          & -14.35          \\
\bottomrule
\end{tabular*}
\end{table}

Propensity score matching can effectively alleviate the endogeneity problem caused by sample selection bias. As shown in Figures 1, 2, the standardised bias of each variable is substantially reduced after matching. and most of the observations were within the common range of values. This result indicates that nearest neighbour matching better balances the differences between the treatment and control groups in the sample, and the assumptions of balance, and common support are satisfied. This paper uses the matched sample for the analysis to estimate the impact of household registration on the employment of the mobile population.

\begin{figure}
	\centering
		\includegraphics[scale=.5]{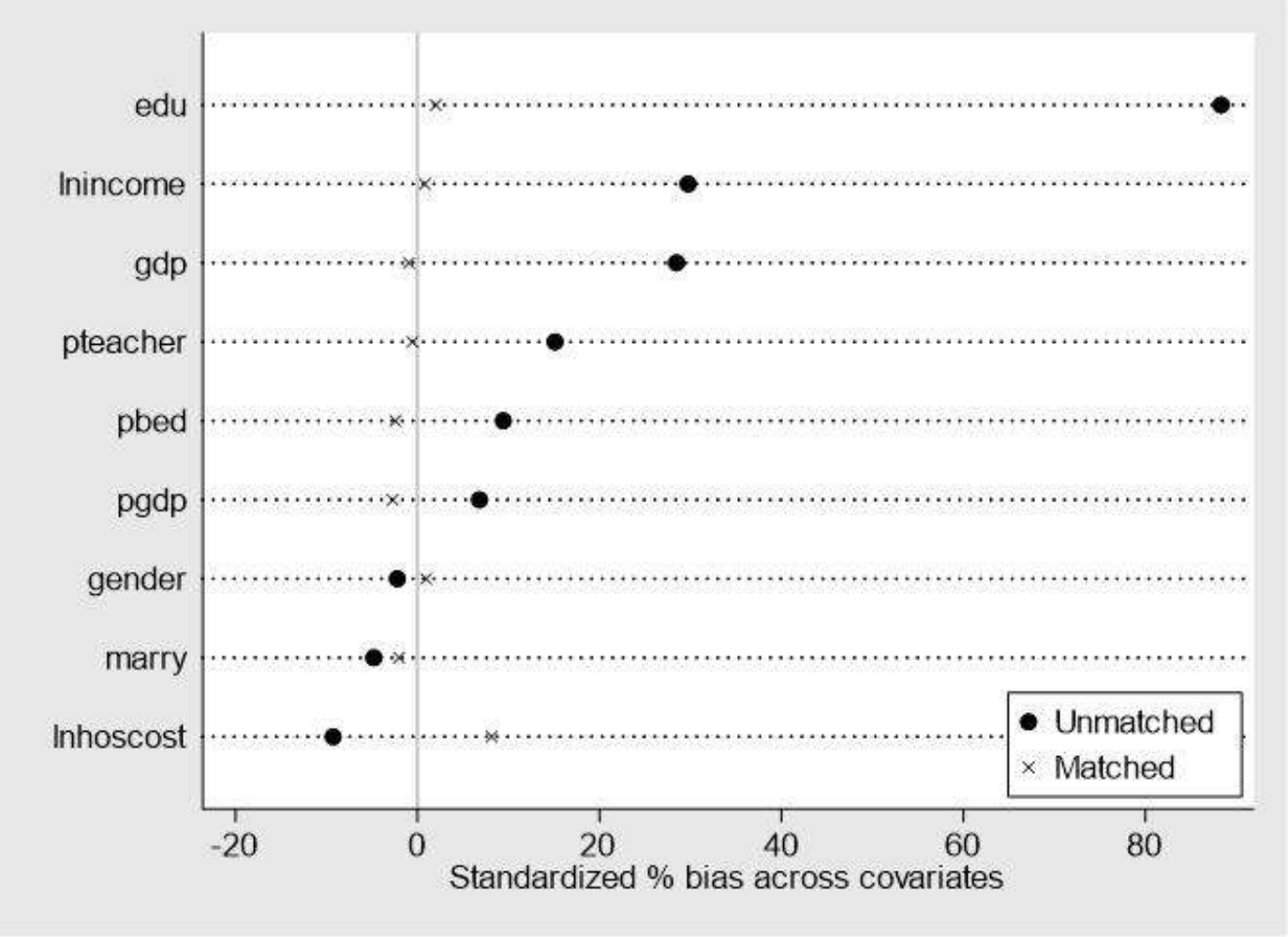}
	\caption{Variable standardisation deviation.}
	\label{FIG:1}
\end{figure}

\begin{figure}
	\centering
		\includegraphics[scale=.5]{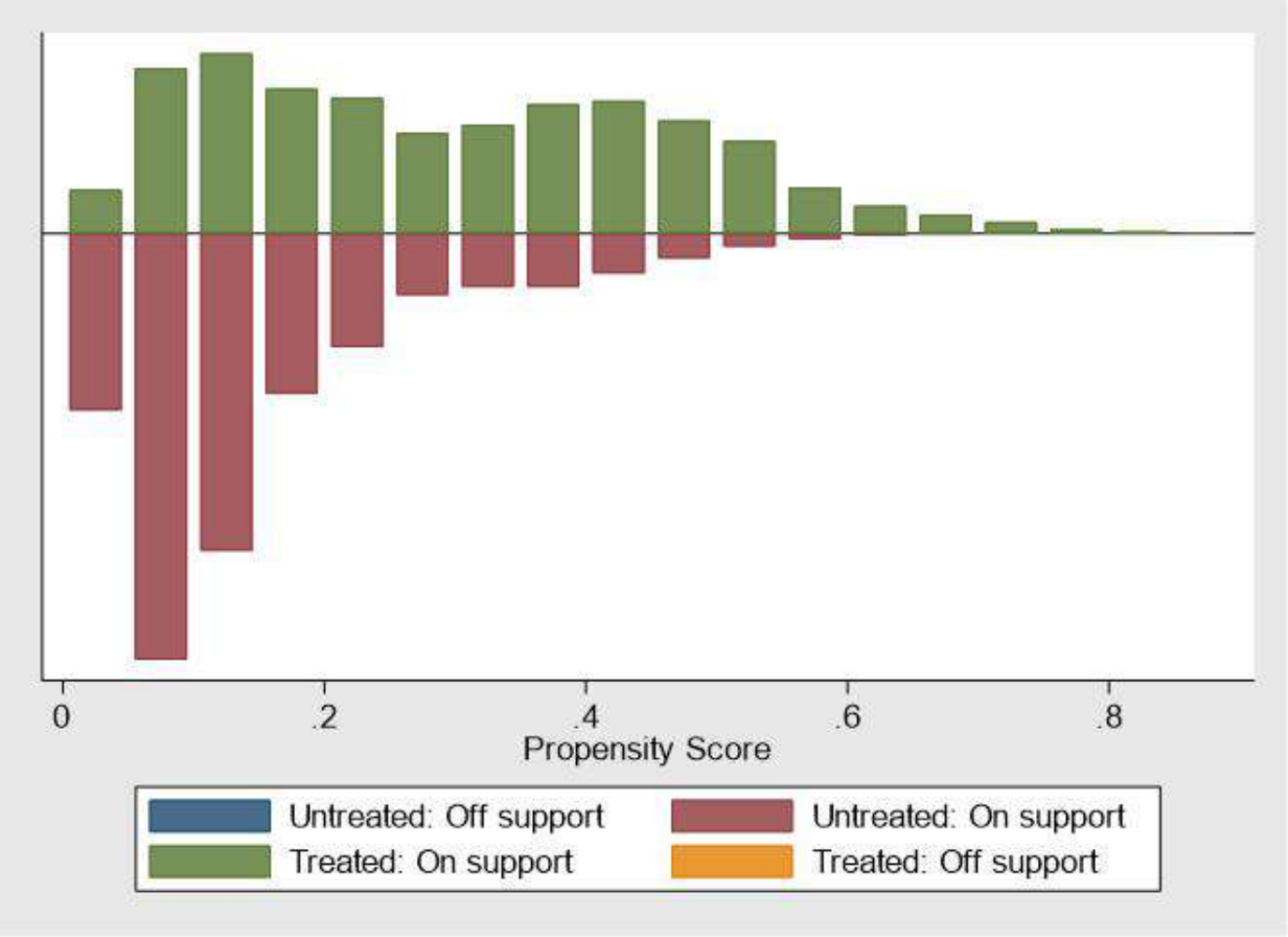}
	\caption{Common range of values for propensity scores.}
	\label{FIG:2}
\end{figure}

Table 6 shows the results of the PSM estimation, where the average treatment effect for all four matching methods is around -0.0678 for the matched sample of individual characteristics control variables, which is the same conclusion as the regression results above: non-agricultural households are less likely to obtain a job compared to agricultural households. Propensity score matching uses likelihood estimation methods by default and does not lend itself to coefficient size comparisons with benchmark regressions based on OLS estimation methods.

\textbf{2.3 Discussion: migrant vs. domiciled population}

\begin{table}[htb]
\caption{Sample selection model. }\label{tbl1}
\begin{tabular*}{\tblwidth}{@{} CCCC@{} }
\toprule
                                      & (13)                     & (14)                  & (15)                       \\
                                      & Mobile population groups & Local household group & Heckman's two-stage return \\
\midrule                                          
hukou                                 & -0.0825 ***              & -0.1387 ***           & -0.0610 ***                \\
                                      & (0.00306 )               & (0.01283)             & (0.01423 )                 \\
\textit{Individual control variables} & $\surd$                  &$\surd$                & $\surd$                    \\
\textit{Urban control variables}      & $\surd$                  & $\surd$               & $\surd$                    \\
\textit{N}                            & 159883                   & 12,000                & 174,225                    \\
R2                                    & 0.097                    & 0.209                 & -                          \\
\midrule    
\textit{LR test}                      & -                        & -                     & P=173.63                   \\
\textit{Lambda}                       & -                        & -                     & -0.1174671                 \\
\bottomrule
\end{tabular*}
\end{table}

Table 7 analyses whether the transmission mechanism of the effect of household registration on employment is affected by the sample's status as " migrant or household population" by using a sample of the household population from eight cities sampled in the 2015 CMDS and a sample of the migrant population from the national sample. Equation 13 shows the regression results for the mobile population sample, while Equation 14 shows the regression results for the domiciled population sample. It can be seen that the employment status of non-agricultural households is significantly lower than that of agricultural households for both the mobile and household populations.

The design of the selection equation is particularly important in the sample selection model. In this paper, logged housing expenditure, logged household income, and years of schooling are used as explanatory variables in the selection equation. In general, the household population has a higher probability of owning a home locally compared to the mobile population, and therefore has a smaller housing expenditure. Similarly, the household population tends to have a greater institutional advantage in terms of schooling and income compared to the mobile population. In the selection equation, all three selection variables are significant at the 1\% level. Also, the likelihood ratio test (chi2 = 173.63) rejected the original hypothesis that the sample selection model should not be used.

Equations 13 and 14 estimate the correlation between household registration and employment status from both the migrant and household population perspectives respectively, and Equation 15 analyses the migrant and household populations as a whole through the sample selection model. According to the final results in Table 7, the coefficients after the sample selection model estimation are reduced and the corrected regression coefficient of household registration on employment of the mobile population is -0.0610. The mechanism of the effect of the nature of household registration on employment in not influenced by the status of the mobile or household population.

\section{Conclusion}

The use of household registration as a control variable for demographics is common in studies on labour economics in China due to the specificity of the dualistic household registration system and the differences in the rights and interests of agricultural and non-agricultural households. In the policy context of family planning, this paper proposes the use of household size and the number of children as instrumental variables for household registration, and discusses qualitatively and statistically validates their relevance and exogeneity. It is finally concluded that the family size and number of children of the mobile population are statistically correlated with the nature of their household registration, in line with the hypothesis of correlation of the instrumental variables. The two instrumental variables passed the over-identification test, proving that the number of children and family size of the migrant population sample have no direct effect on their willingness to settle in the city upon re-entry. The hypothesis of exogeneity is met when family size and the number of children of the migrant population are used as instrumental variables for household registration.

Starting from the institutional context of China's dualistic household registration policy, this paper empirically analyses the impact of the nature of household registration on the citizenship of the migrant population through a two-stage least squares approach. After controlling for urban and individual control variables as well as fixed effects, the following conclusions are drawn: non-agricultural hukou has about 20.2\% lower willingness to settle and about 7.28\% lower employment level in inflowing cities compared to the agricultural hukou group; the mechanism of the effect of the nature of hukou on employment in not influenced by the status of the mobile or hukou population. Whether analysed in terms of subjective willingness to settle or objective employment, agricultural hukou shows a higher level of citizenship compared to non-agricultural hukou.

Although the division between agricultural and non-agricultural hukou has been phased out of China's hukou management system since 2014, the vast differences in the rights and benefits attached to the two types of hukou and the decades-long duration of their implementation still have a profound impact on the regional allocation of labour endowments in China. The nature of household registration remains a very important control variable in the study of issues related to labour economics in China. The main contribution of this paper is to propose an instrumental variable for the household registration system and to argue for the feasibility of this instrumental variable.

\bibliography{cas-refs}

\end{document}